\documentclass[10pt, conference]{IEEEtran}
\IEEEoverridecommandlockouts
\usepackage{cite}
\usepackage{lmodern}
\usepackage{soul}
\usepackage{amsmath,amssymb,amsfonts}
\usepackage{algorithmic}
\usepackage{graphicx}
\usepackage{textcomp}
\usepackage{xcolor}
\usepackage{xspace}
\usepackage{dirtytalk} %
\usepackage{enumitem} %
\usepackage{url}
\usepackage{lineno}
\usepackage{tabularx}
\usepackage{booktabs}
\usepackage{threeparttable}
\usepackage{sparklines}
\usepackage[capitalise,noabbrev]{cleveref}

\newcommand{\name}{\textsc{FixAlly}\xspace}
 
\def\BibTeX{{\rm B\kern-.05em{\sc i\kern-.025em b}\kern-.08em
    T\kern-.1667em\lower.7ex\hbox{E}\kern-.125emX}}
\begin{document}

\title{Automated Code Fix Suggestions for\\ Accessibility Issues in Mobile Apps}

\author{\IEEEauthorblockN{Forough Mehralian}
\IEEEauthorblockA{
\textit{Apple}\\
Seattle, USA \\
mehralian@apple.com}
\and
\IEEEauthorblockN{Titus Barik}
\IEEEauthorblockA{
\textit{Apple}\\
Seattle, USA \\
tbarik@apple.com}
\and
\IEEEauthorblockN{Jeff Nichols}
\IEEEauthorblockA{
\textit{Apple}\\
Seattle, USA \\
jwnichols@apple.com}
\and
\IEEEauthorblockN{Amanda Swearngin}
\IEEEauthorblockA{
\textit{Apple}\\
Seattle, USA \\
aswearngin@apple.com}
}

\maketitle

\begin{abstract}

Accessibility is crucial for inclusive app usability, yet developers often struggle to identify and fix app accessibility issues due to a lack of awareness, expertise, and inadequate tools. Current accessibility testing tools can identify accessibility issues but may not always provide guidance on how to address them. We introduce \name, an automated tool designed to suggest source code fixes for accessibility issues detected by automated accessibility scanners. \name employs a multi-agent LLM architecture to generate fix strategies, localize issues within the source code, and propose code modification suggestions to fix the accessibility issue. Our empirical study demonstrates \name's capability in suggesting fixes that resolve issues found by accessibility scanners---with an effectiveness of 77\% in generating plausible fix suggestions---and our survey of 12 iOS developers finds they would be willing to accept 69.4\% of evaluated fix suggestions.  
\end{abstract}


\section{Introduction}
\label{Sec:Introduction}
The increasing reliance on mobile apps for everyday tasks underscores the necessity of ensuring accessibility for all. Despite the existence of guidelines aimed at assisting developers in creating more accessible apps~\cite{android_guide, ios_guide, wcag}, research shows that many apps are still released with numerous accessibility issues~\cite{alshayban2020accessibility, vendome2019can, zhang2021screen, ross2018examining}. Developers often struggle with building accessible apps because they  lack awareness of accessibility requirements~\cite{alshayban2020accessibility} or have limited knowledge and expertise in effectively addressing accessibility issues~\cite{bi2022accessibility}.

Existing accessibility scanning tools---such as Accessibility Scanner~\cite{googleAccessibilityScanner} for Android and Accessibility Inspector~\cite{accessibilityInspector} for iOS---helpfully verify compliance of each app screen with rules derived from accessibility guidelines. In addition to these rule-based techniques, some automated tools dynamically examine apps using assistive technologies to detect issues that manifest during real-time interactions~\cite{mehralian2022oversight, salehnamadi2022groundhog}. However, current tools provide insufficient support for maintaining app accessibility~\cite{bi2022accessibility} because fixing the large number of issues reported by these tools remains a significant challenge. While single-issue fix techniques address problems like color issues~\cite{zhang2023automated}, missing labels~\cite{mehralian2021data}, text scaling problems~\cite{alotaibi23icsme}, and touch target size~\cite{alotaibi2021automated}, these single-issue fix techniques have notable limitations. According to documentation, Accessibility Inspector reports 7 categories of issues~\cite{accessibilityInspector}, and these single-purpose approaches can fix only a small subset.

To bridge this gap in tooling between single-purpose fix approaches and source code, we investigate an automated \emph{plan-localize-fix} technique---implemented as a tool called \name---to fix various types of accessibility issues reported by scanners such as the Accessibility Inspector~\cite{accessibilityInspector}. To understand the challenges of fixing issues detected by accessibility scanners in source code, we first conducted formative interviews with five developers. Our developers indicated that: 1) multiple strategies can address a single issue, 2) appropriate fixes must consider not only accessibility guidelines, but also the integrity of the app's design and functionality, 3) implementing a fix frequently requires modifications beyond the problematic element, and 4) identifying these relevant locations in the code to apply fixes is the most time-consuming step.

To address these needs, \name employs a multi-agent LLM architecture capable of proposing \emph{plausible} fix suggestions for issues reported by an accessibility scanner. In this context, a plausible fix is defined as a modification that passes the accessibility checks of the automated scanner for the target issue without introducing new ones or removing functionality. \name analyzes an open-source mobile app to detect various accessibility issues. \name localizes issues within the source code and proposes fix suggestions to resolve the issue using a suggestion generation engine. Each proposed fix suggestion aims to resolve the targeted accessibility issue without introducing new ones or compromising app functionalities. \name also assists the developer in the decision-making process to select the strategy that best aligns with the app's design and requirements.

The contributions of this paper are:

\begin{itemize}
    \item A novel plan-localize-fix technique---operationalized as an automated tool using a multi-agent LLM architecture---that generates code suggestions to fix accessibility issues in mobile apps.
    \item An empirical evaluation  on 205 issues from 14 iOS apps built using SwiftUI, a declarative framework that allows developers to define the desired UI attributes and behavior~\cite{swiftui}. Our evaluation demonstrates an effectiveness of 77\% of \name in proposing plausible fix suggestions for accessibility issues. 
    \item A survey of 12 iOS app developers, finding the tool was most helpful for less experienced developers in allowing them to explore multiple solutions when resolving accessibility bugs. Even experienced developers found it helpful that \name  localized the issue in the code. 
\end{itemize}

\begin{figure}[t]
  \includegraphics[width=\columnwidth]{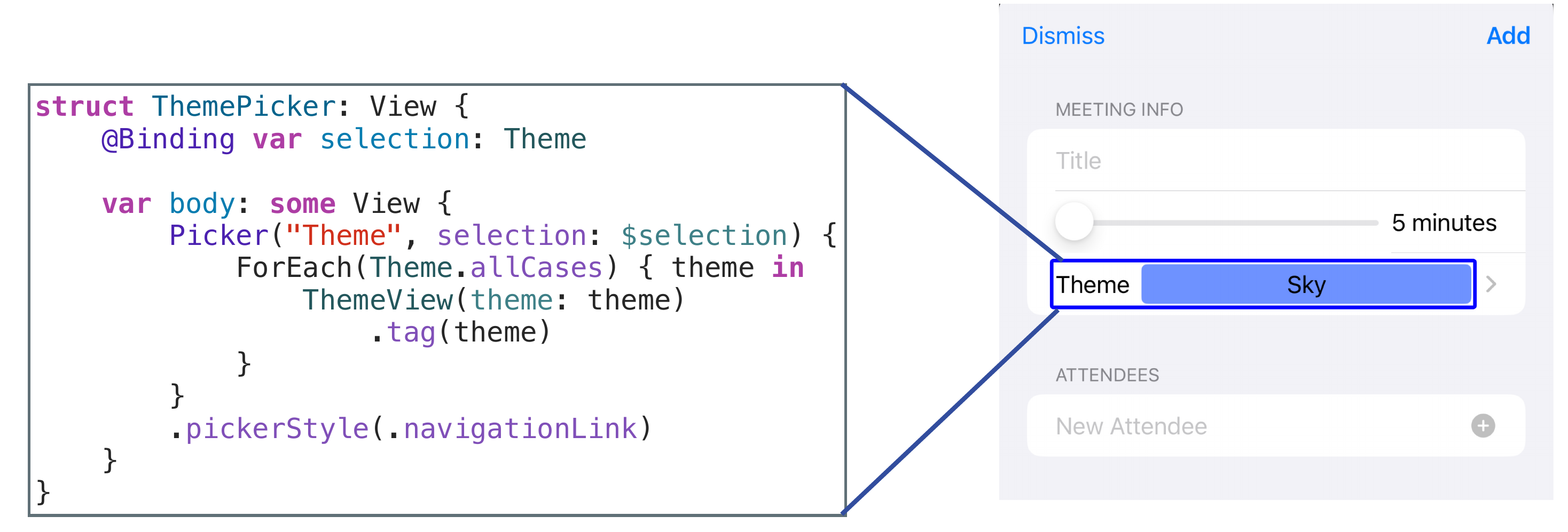}
  \caption{Implementation of a dropdown list in SwiftUI.}
  \label{fig:background}
\end{figure}

\begin{figure}[t]
  \includegraphics[width=\columnwidth]{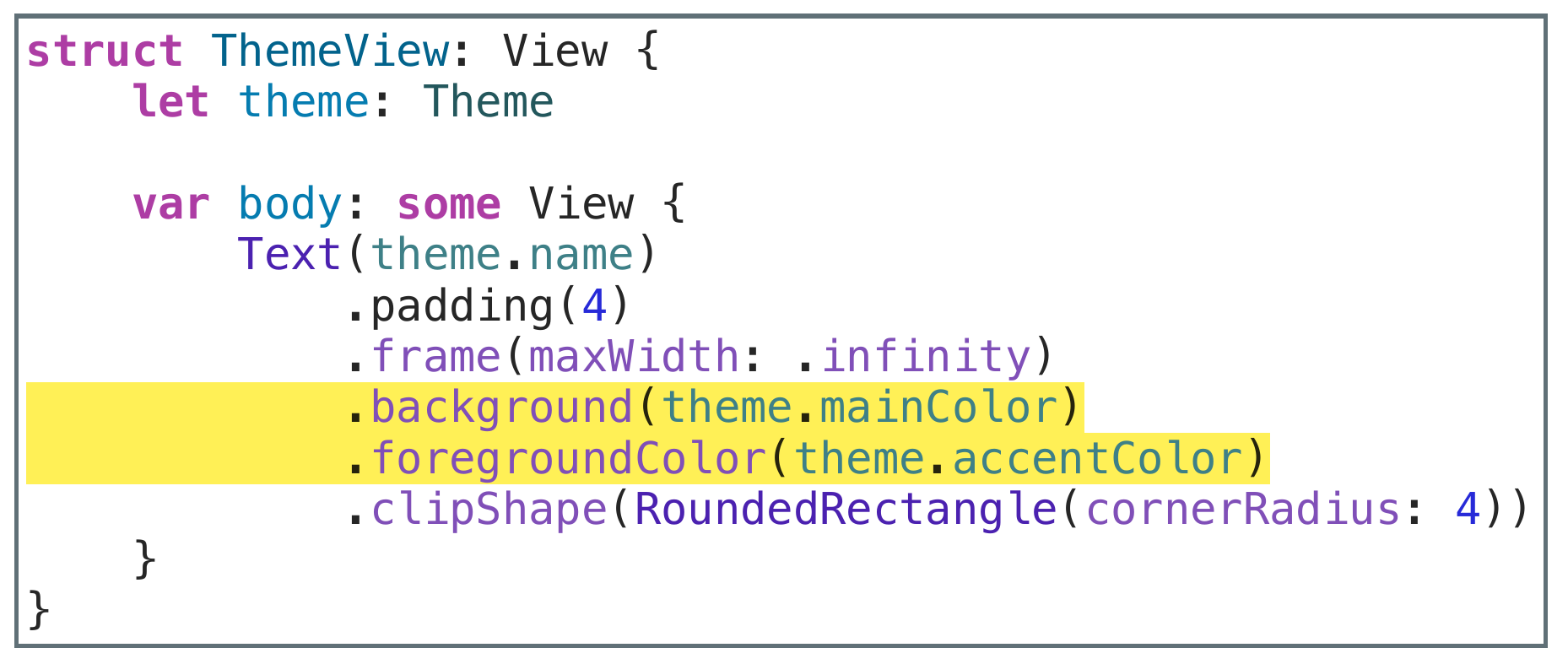}
  \caption{Localization of the color contrast issue in the source code.}
  \label{fig:background2}
\end{figure}

\section{Background: Comparing Android and iOS}
\label{Sec:Background}
Many of the existing approaches focus on Android applications which require different techniques than iOS applications. 
Declarative programming languages, such as SwiftUI, represent a contemporary paradigm for building GUIs in mobile apps by enabling developers to define the desired UI and its behavior using concise syntax. Figure~\ref{fig:background} illustrates how a dropdown list is implemented in SwiftUI. This approach contrasts with traditional imperative methods, where developers must meticulously specify attributes of each UI element and manage its state. For example, in Android, the UI specification is defined using an XML file, with Java classes binding behavior to each element. Developers can also modify UI attributes dynamically through specified behavior in declarative languages. Therefore, properly locating GUI problems in the project cannot be effectively achieved through analysis or modification of the static GUI specification alone. The dynamic specification of UI attributes, which is the core of declarative programming languages, introduces additional challenges in identifying GUI problems, such as accessibility issues, within the source code. 

Consider the color contrast failure issue for the annotated dropdown list in Figure~\ref{fig:background}, as reported by Accessibility Inspector. Developers typically employ various strategies to address such issues, such as adjusting background colors, modifying text colors, or increasing font sizes. However, these strategies cannot be applied in the provided code snippet that demonstrates the element's implementation. Instead, the appropriate place to adjust the color for this element is within the ThemeView called from the ThemePicker (Figure~\ref{fig:background2}). Locating this correct position necessitates navigating through the UI hierarchy, understanding the semantics of the GUI, and comprehending the structure of the source code required for implementing these elements. This complexity has made localization a challenging task.

\section{Formative Study}
\label{Sec:Formative}
To elicit the process developers follow to fix accessibility bugs found by an accessibility scanner in iOS apps, we conducted formative interviews with five iOS app developers in our company. The developers had at least 1 year of experience in developing SwiftUI apps and median of intermediate accessibility familiarity (1 -- No Experience to 5 -- Expert). During the study, the developers used the Accessibility Inspector and Xcode to detect issues in the Landmarks app\footnotemark\hspace{0.5em}running on an iOS simulator. We asked them to think-aloud while they detected and fixed as many issues as they could within 1 hour. At the end, we asked them follow-up questions about their experiences and their ideas on any tools that could improve their process in finding and fixing the accessibility issues. All sessions took place virtually over Webex. Some developers built and tested the app locally while sharing their screen and some developers remotely controlled the screen of the lead researcher who also had the app and simulator built and running in Xcode. A second researcher observed and took notes for four out of five sessions. We recorded audio, video, and notes for each session. 
\footnotetext{https://github.com/pd95/SwiftUI-Landmarks}

To analyze the data, we annotated the transcripts and built an affinity diagram~\cite{affinitydiagram} where one paper author led the annotation and initial grouping, and another author read and also validated the themes. 

\subsubsection{Accessibility Bug Fixing Phases}
We examined the developers' overall process in fixing the accessibility bugs with two goals: first,  understanding the varied activities (e.g., localizing an issue, fixing it in the code) within the whole process, and second, understanding inefficiencies within each of those activities where an automated tool could help. We found that developers accessibility bug fixing workflow can be grouped into the following phases: \textit{hypothesis formation \& fix planning}, \textit{localization}, and \textit{code editing and validation}. We ultimately designed the architecture of \name around these three phases. 

\textit{Hypothesis formation and fix planning:}
All developers in our study had at least intermediate accessibility knowledge and could understand the issues reported by Accessibility Inspector. They were all able to propose hypotheses to diagnose one or more issues, and come up with a high level plan to fix one or more of them. In some cases, developers directly proposed a fix plan for some issues because they were already highly familiar with the issue and could quickly come up with the fix. In other cases, developers mentioned multiple hypotheses that the issue could be related to, and validated which to test after localizing the impacted UI element in the source code. 

\textit{Localization:} The output of Accessibility Inspector, and typically other accessibility scanners, is a screenshot highlighting the impacted UI element and the ability to highlight the element with the issue on a live device. While these tools provide some metadata and inspectors to examine the running app's hierarchy, they provide little help with localizing UI elements in code. In our study, developers used a variety of methods to localize UI elements in the code including searching for specific text strings found in the interface. Others tried to match the visual hierarchy in the interface with the hierarchy of views in the source files by looking at them one by one. Sometimes they also looked for specific SwiftUI modifiers in the code based on their hypothesis, or examined the view hierarchy in the accessibility inspector for class names or metadata they could search for. Localization was where they predominantly spent the most time during the study. 

\textit{Code editing and validation:}
After fix planning and localization, developers applied their fix to the code. They then re-ran the Accessibility Inspector audit on the same screen again to confirm the issue was resolved. If the issue was not resolved, they could repeat the overall process as long as time allowed. 

Developers did not necessarily complete the bug fixing phases in the same order. In all cases, developers either localized or came up with a hypothesis or fix plan first. Some developers first attempted to localize a UI element in the source code before coming up with a hypothesis or fix plan, while others first described a hypothesis or fix plan before attempting to localize the UI element. Some developers repeated this process multiple times before confirming the fix by validating it no longer appeared in the Accessibility Inspector.  

\subsubsection{Design Goals}   
At the closing of the session, we asked the developers to describe their overall process for fixing the accessibility issues, and to provide feedback on any tools that could expedite their process. We then formulated list of design goals that we incorporated into our system. The design goals were motivated by the challenges developers faced throughout the accessibility bux fixing process and suggestions they provided to address these challenges. 

\textit{Design Goal 1 -- Localize impacted UI elements in source, and automatically apply fix suggestions:}
Developers struggled the most during our interviews with finding impacted UI elements in source code, which is often the starting point for making a fix. This added tedious, and unnecessary friction to their bug fixing process. 
Developers utilized different heuristics to localize the issue, including searching for textual elements, mapping their mental model of elements in the screenshots to the UI structure in the code, looking for specific accessibility attributes they were planning to modify, or combinations of these approaches.
Thus a key goal of \name is to automatically localize potential fix locations in code for detected issues to remove this inefficiency and also generate fix suggestions in the form of patches so they can be applied automatically.

\textit{Design Goal 2 -- Provide multiple fix suggestions for an issue:}
Some developers struggled to come up with hypothesis for fixes, and even if they had a plausible hypothesis, some did not know how to make the corresponding code changes to test it. Furthermore, some issues often can be fixed in multiple ways. For example, contrast issues can be fixed by increasing text font size, or changing the color of UI elements or background. Developers noted other considerations and potential side effects in choosing a correct fix including consulting with design teams, support for other languages, and impact on the layout of other areas of the screen or the application. Some developers also struggled to understand some issues reported by Accessibility Inspector, and requested better suggestions more contextualized and specific to their code. To help developers in understanding each suggested fix, \name also includes information about the model's fix plan and reasoning along with each generated fix suggestion.

\begin{figure*}
\centering
  \includegraphics[width=\linewidth]{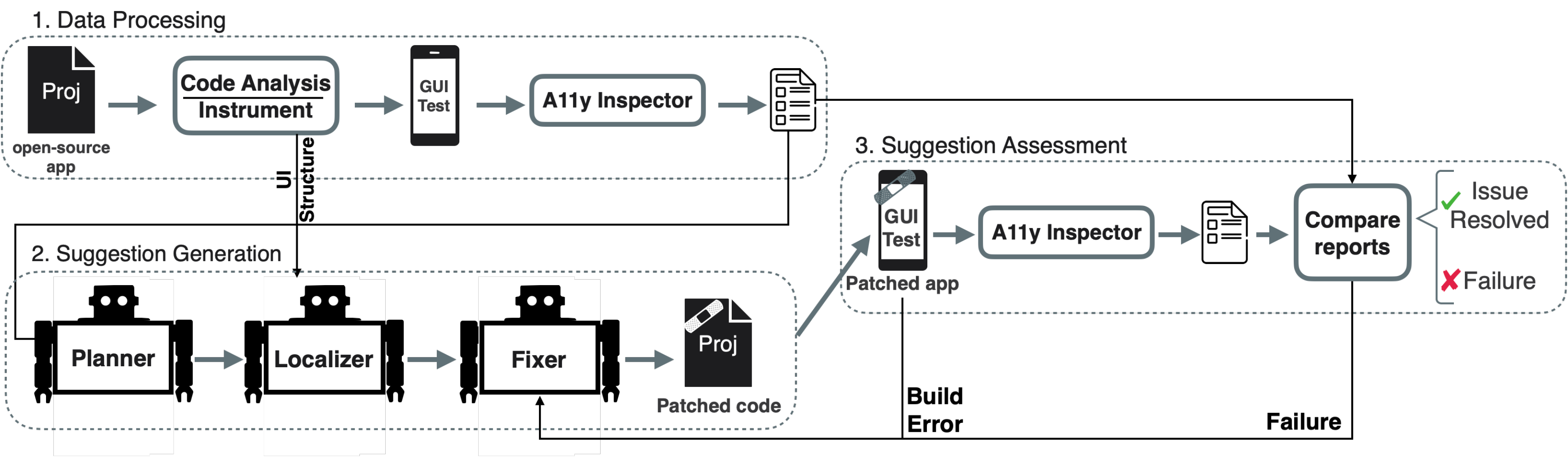}
  \caption{\name's approach, consisting of 1) \textit{Data Processing} which instruments the application and navigates to various screens via GUI tests to capture accessibility scans and screenshots, 2) \textit{Suggestion Generation} which uses a multi-agent LLM architecture to generate fix suggestions for each detected issue from the input screenshot, source code, and issue descriptions, and 3) \textit{Suggestion Assessment} which captures a new accessibility scan of the patched app and GUI screen and compares it to the prior report to determine if the fix suggestion resolved the issue. \label{fig:approach}}
\end{figure*}

\newcommand{\prep}{Data Processing\xspace}
\newcommand{\engine}{Suggestion Generation Engine\xspace}
\newcommand{\assess}{Suggestion Assessment\xspace}

\section{Approach}
\label{Sec:Approach}
\cref{fig:approach} shows an overview of \name, with its three main modules: \prep, \engine, and \assess.

\name takes as input an app with GUI tests, each designed to navigate to different screens of the app. The XCTest framework~\cite{xctest}, integrated into Xcode IDE, allows iOS developers to create automated test scenarios that navigate the app to the screens targeted for accessibility assessment. Additionally, existing automated app crawlers can help developers generate these test scenarios automatically. These crawlers either perform random actions on the screen~\cite{xcmonkey} or analyze UI elements to systematically explore all possible actions on those elements~\cite{su2017guided}.

Automated testing frameworks recommend that developers use unique identifiers for each element, serving as a bridge between testing frameworks and apps. To ensure every UI element is associated with an identifier, the \prep module statically analyzes the source code, restores the UI hierarchy, and instruments the app to insert a unique accessibility identifier for each UI element~\cite{accessibilityIdentifier} in the input SwiftUI app. To find accessibility issues, the \prep module runs GUI tests to navigate to varied screens in the app. It then uses Accessibility Inspector~\cite{accessibilityInspector} to obtain a report of  accessibility issues on an app screen navigated to by a GUI test. The output of the scanner includes a description of each issue, the identifier of the impacted UI element or its parent elements in the UI hierarchy, and a screenshot of the app with the location of the problematic element.
The \engine processes the information for each issue to generate multiple fix suggestions for addressing the issue, and patched project to verify whether suggestions resolve the issue (Details in Section~\ref{Sec:PatchGen}). Inspired by \textit{Design Goal 1}, the \engine is capable of localizing the source of issues among the code for an app across multiple files, and generating patches that developers can automatically apply to fix the issue. 

To verify that generated fix suggestions can plausibly fix each issue, \name evaluates each code modification in its \assess module. This module takes the modified code snippet to generate a patched version of the project, attempts to build the app, and runs the same GUI test to navigate to the target screen where the issue was detected. It then uses the Accessibility Inspector to audit the screen and compare the report with the initial report.

\engine fails when the generated fix has build errors, was not able to resolve the accessibility issue, or introduced new issues. The model may also inadvertently comment out sections of code or remove necessary screen elements. \name feeds these failure messages back to the \engine to self-reflect and let it revise the modified code snippet. This iterative process mirrors a developer’s approach of assessing and revising modifications in response to issues. In our work, we configure the number of iterations for this feedback loop, setting it to 3 to allow sufficient opportunities for the model to refine its fix suggestions. Finally, if the fix suggestion successfully resolves the issue without introducing other issues or removing functionalities, the \assess module marks it as a fix suggestion that can be shown to developers. \name provides developers with multiple fix suggestions for an issue inspired by \textit{Design Goal 2}.

\begin{figure}
\centering
  \includegraphics[width=0.65\linewidth]{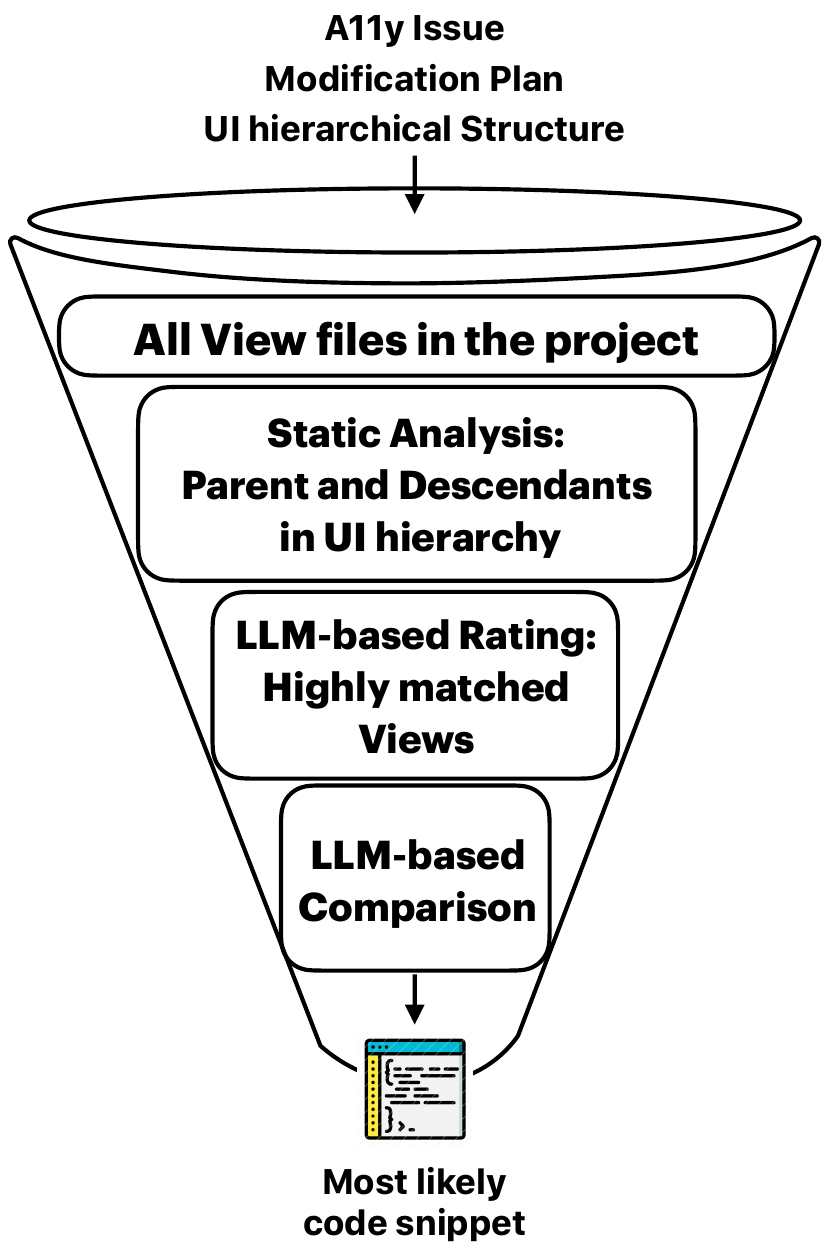}
  \caption{Multi-level, hybrid localization architecture: Static analysis of the UI hierarchy identifies parent and descendant views. LLM-based rating evaluates the match of each individual code snippet to the screenshot. Finally, LLM-based comparison examines the highly matched views to determine the most likely code snippet for applying the fix.}
  \label{fig:localizer}
\end{figure}

\begin{figure*}
    \centering
    \includegraphics[width=\linewidth]{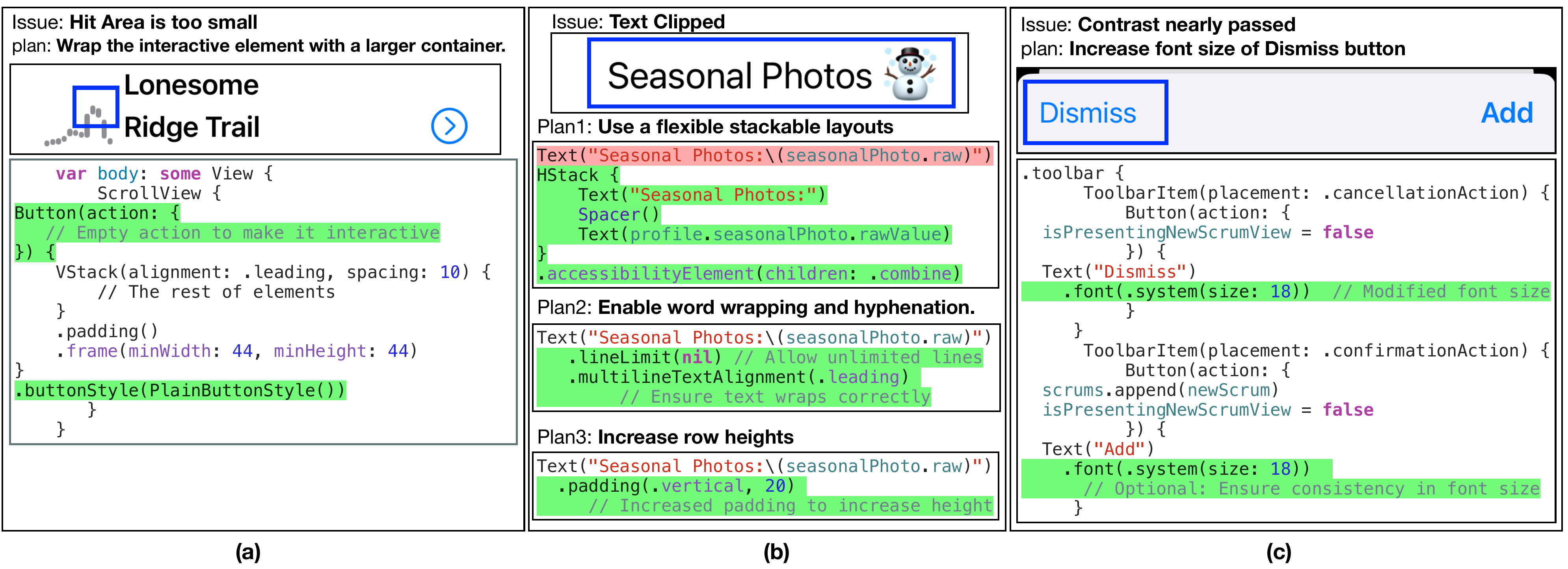}
    \caption{Three different types of issues that were fixed by \name. (a) The tool addresses the issue of small hit target size by identifying a group of semantically related elements and merging them into an interactive container. (b) Shows three generated fix suggestions for Text Clipped issue. The first plan involves replacing a single element with a group of elements while preserving the functionality(c) \name addresses the contrast issue of the ``Dismiss'' button, while also maintaining design integrity by applying the font change to another similar button.}
    \label{fig:results}
\end{figure*}

\section{Suggestion Generation Engine}
\label{Sec:PatchGen}
The Suggestion Generation Engine (Figure~\ref{fig:approach}.2) is suggests fix strategies for each reported accessibility issue. Figure~\ref{fig:background} shows the \engine, illustrating three agents in this module: Planner, Localizer, Fixer. These agents are responsible for performing specific steps that developers take in fixing an accessibility bug, as we observed in our Formative Study (\cref{Sec:Formative}). The details of each agent are below. For each agent, we currently use GPT-4o~\cite{gpt4} for the LLM model. The specifics of the prompt, along with the source code, are available in the supplementary material submitted with the paper.

\subsection{Planner: Suggesting fix plans}
Each accessibility issue may be resolved in various ways. According to our Formative Study~\ref{Sec:Formative}, developers prefer solutions that not only fix the issue without introducing new issues, but also align with the design decisions of the app and work well in different modes, such as dark mode or horizontal mode. Providing different options for fixing an issue allows the developer to choose the one that best fits the overall design and functionality. The Planner agent facilitates this by generating natural language suggestions of strategies to fix each issue. For example, agent may suggest a plan of ``adjust background color for better contrast'' to fix a ``contrast failed'' issue for a UI element. 

\name provides the Planner agent with an annotated screenshot of the app along with the issue description from the Accessibility Inspector. Its task is to identify the most relevant accessibility guideline related to the reported issue and list techniques to resolve it in natural language. \name aims to leverage the language model's knowledge of accessibility fixes across various platforms, such as the web, to provide suggestions that can be adapted for mobile apps on specific platforms like iOS. \name instructs the agent to avoid suggesting solutions that are not applicable to the source code, filtering out recommendations such as using third-party tools to test the app. In our evaluation, \name instructs the Planner to return three alternative plans for each issue, though this number can be configured.

\subsection{Localizer: Finding the relevant code snippet}
Localizing an issue involves identifying the precise location in the source code where the fix plan should be applied. Source code projects often span numerous files with thousands of lines, potentially exceeding the token limits that language models can effectively query in one request or retain in their context window. To manage this complexity, we employ a multi-level localization approach using static code analysis and LLM code analysis. Figure~\ref{fig:localizer} illustrates the different levels of localization, including one static code analysis step and two LLM-based steps for issue-to-code mapping.

First, the Localizer identifies all of the View files of the project, identifiable by the `\texttt{import SwiftUI}' statement and structures extending the View class. Then, it extracts candidate code snippets from these view files through static code analysis. The key insight that the Localizer uses to optimize this process is that the behavior and design of UI elements are predominantly influenced by their descendants or ascendants. The Static Analysis module takes the accessibility identifier of the problematic element and traverses the pre-analyzed UI structure to return all descendant views and the parent of that element to form a set of candidate code snippets. \name adds these accessibility identifiers to each view in the code in the \prep phase, where it also captures the UI hierarchy (Recall \cref{Sec:Approach}).

Next, the Localizer generates an LLM-based rating for the code snippets filtered by the static analysis module. It matches each snippet to the modification plan for the issue and the screenshot highlighting the problem. Due to the limited context window of LLMs, it may not be feasible to consider and compare all candidate snippets simultaneously to find the correct one to apply the fix. Instead, the Localizer agent first assesses each snippet individually based on detailed issue information, the screenshot, and the proposed fix plan to determine its suitability. \name instructs the agent to map the problematic element and other elements in its vicinity to the source code, consider the fix plan, and rate the likelihood that the snippet is the correct location. This approach mirrors various techniques developers use to localize accessibility issues in the source code, as mentioned in \cref{Sec:Formative}. The language model's code comprehension capabilities, combined with its knowledge of accessibility, enable it to rate the alignment of the code with the highlighted element in the screenshot and the fix strategy. \name selects the code snippets with the highest match rates for further comparison by the Localizer agent. The Localizer agent ranks the highly rated snippets and selects the one most likely to be the correct location for applying the fix.

\subsection{Fixer: Modifying the code}
The goal of the Fixer agent is to apply the fix strategy to the selected code snippet. The Fixer agent receives the issue details, the fix strategy for the issue, and the most relevant code snippet to be modified based on the plan. Instead of relying on the agent to generate a diff, we instruct the agent to directly apply the modifications and return the updated code. This approach avoids potential inaccuracies in diff generation by language models, which may struggle with accurately calculating changes across lines of code and managing the required number of tokens. Finally, the \engine module applies the diff and creates a new copy of the project. 

\subsection{Output}
It also stores a diff of the updated code and the original buggy code to provide a visualization of the code suggestion for developers to examine. Figure~\ref{fig:results} shows examples of fix suggestions generated by \name, demonstrating its ability to address different types of issues and apply different strategies to fix an issue. The output of \name is currently a diff visualization of the code for each fix suggestion along with the fix plan and the model's explanation of the changes. 

As shown in Figure~\ref{fig:results}(a), \name correctly identifies a group of semantically related elements and merges them into an interactive container to address the issue of small hit target size. In contrast, Figure~\ref{fig:results}(b) demonstrates the capability of \name in splitting a single element into a group of elements to address the Text Clipped issue. Figure~\ref{fig:results}(c) shows that when fixing the contrast issue of the ``Dismiss'' button, the tool also maintains the app's design integrity by applying the proposed fix to a similar ``Add'' button in the toolbar, demonstrating its ability to fix accessibility issues while maintaining consistency between similar UI elements.

\section{Evaluation}
\label{Sec:Evaluation}
We evaluated \name through the following research questions:
\begin{enumerate}[leftmargin=1.1cm,label=\bfseries RQ\arabic*.]
    \item (Effectiveness) How effective is \name in generating code fixes for accessibility issues detected by an accessibilty scanner?
    \item (Efficiency) What is the efficiency of \name in terms of time, the number of attempts, and the cost? 
    \item (Helpfulness) How helpful are the proposed fixes for developers?
\end{enumerate}
\subsection{Experimental Setup}
We evaluated our approach using 14 open-source apps sourced from GitHub. Specifically, we randomly selected apps from two GitHub repositories that catalog open-source iOS apps~\cite{awesome-ios, open-source-ios-apps}, excluding apps not built using SwiftUI. For each app, one of the authors attempted to build the app successfully within a 30 minute window. We also excluded apps with build errors due to dependencies, external packages, or very old iOS versions from the dataset. \cref{tab:RQ1} provides a list of the apps included in our study. The list of apps with their corresponding GitHub links is also available in our supplementary materials.
\name's implementation leverages LLM agents based on GPT-4o, which features a 128K context window and has a knowledge cut-off date of October 2023. We conducted the experiments on a MacBook M1 Pro equipped with 32GB of RAM, a typical computer setup for development. We used Xcode 15.0, the latest available version, to build the apps, and we installed and tested them on an iPhone 12. 

\subsection{RQ1. \name's effectiveness}
We assessed the efficacy of \name by evaluating its ability to propose fix suggestions for 204 issues across 22 screens of SwiftUI iOS apps. \cref{tab:RQ1} presents the outcomes of \name in generating fix suggestions. We use the term \emph{plausible} to indicate that the generated fix resolved the targeted issue while maintaining app functionality and without introducing new issues. 

\begin{table}
    \centering
    \caption{\name's effectiveness in generating fixes \\for accessibility issues\label{tab:RQ1}}
\begin{threeparttable}
\begin{tabular}{lrrr}
\toprule
\textbf{App}\tnote{1}   & \textbf{Screens} & \boldmath$n$ & \textbf{Plausible Fix (PF)} \\
\midrule
ARPlasticOcean & 1       & 5     & 4             \\
Calculator     & 1       & 24    & 22            \\
DeTeXt         & 1       & 1     & 1             \\
DesignRemakes  & 1       & 1     & 0             \\
ExpenseTracker & 1       & 5     & 4             \\
Fingerspelling & 1       & 4     & 3             \\
Instagram      & 2       & 5     & 3             \\
Landmarks      & 3       & 55    & 48            \\
Ratio         & 1       & 15    & 10            \\
Scrumdinger    & 2       & 7     & 6             \\
Go Cycling     & 4       & 74     & 52              \\
DesignCode    & 1       & 8     & 7             \\
GradeCalc     & 2       & 7     & 3             \\
Sunshine     & 1       & 1     & 1             \\
\midrule
Total          & 22      & 205   & 158           \\ 
\bottomrule
\end{tabular}
\begin{tablenotes}
\item[1] Open-source apps from GitHub.
\end{tablenotes}
\end{threeparttable}
\end{table}

\name demonstrated a 77\% effectiveness in automatically generating fix suggestions for accessibility issues, where effectiveness means it successfully produced at least one plausible suggestion out of three suggestions for 157 out of 204 issues. Furthermore, for 129 (63\%) of these issues, \name generated two or three plausible fix suggestions, providing developers multiple options to consider.  

We also assessed the categories of issues that \name can generate fix suggestions for. Our dataset contains nine different types of issues as shown in \cref{tab:RQ1-1}. According to the documentation for Accessibility Inspector~\cite{AppleAccessibilityAudits}, these issues encompass the categories of Element description, Element detection, Hit region, Contrast, Clipped text, Traits, and Dynamic type. However, our dataset does not contain trait issues: we found that even when modifying some apps to purposefully contain these issues, Accessibility Inspector detected these only on the iOS simulator and not on the physical device used for our experiments. Excluding Traits, \name could successfully resolve at least one issue from each type.

\begin{table}
    \caption{\name's effectiveness in fixing different types of issues\label{tab:RQ1-1}}
    \centering

\begin{threeparttable}
\begin{tabularx}{\columnwidth}{lXrr}
\toprule
\textbf{Category} & \textbf{Issue type} & \boldmath$n$ & \textbf{PF}\tnote{1}\\
\midrule
Clipped text & Text clipped                                      & 15     & 11  \\
\addlinespace
Contrast & Contrast failed                                   & 27    & 21  \\
Contrast & Contrast nearly passed                            & 21     & 17  \\
\addlinespace
Dynamic type & Dynamic Type font sizes are partially unsupported & 16     & 11  \\
Dynamic type & Dynamic Type font sizes are unsupported           & 58    & 39 \\
\addlinespace
Element description & Element has no description                        & 43    & 38 \\
Element description & Label not human-readable                          & 3     & 3  \\
Element detection & Potentially inaccessible text                          & 8     & 7  \\
\addlinespace
Hit region & Hit area is too small                             & 21    & 17 \\
\bottomrule
\end{tabularx}
\begin{tablenotes}
\item[1] Indicates the number of issues with at least one plausible fix.
\end{tablenotes}
\end{threeparttable}

\end{table}

To understand the failures of \name, the first author manually inspected a subset of the generated fix suggestions that did not resolve the accessibility issues. Their analysis suggests that these failures may stem from shortcomings in the planning, localization, or fixing phases. For example, for the issue ``Text Clipped'' only one of the fix suggestions was plausible. Two out of three plans generated by the planner were irrelevant, indicating that not all issues may have multiple plausible solutions. Additionally, in some cases, the tool may select incorrect code snippets for fixing. When there is insufficient information about the problematic element, the model might fail to identify the relevant code snippet, leading to ineffective fix suggestions. Even if localization is accurate, the generated code may contain build errors or be ineffective. Despite these issues, the proposed plans, related code snippets, and SwiftUI accessibility attributes generated by the tool can still help developers devise a fix more quickly. Furthermore, some reported failures were due to false positives from the Accessibility Inspector. For instance, after testing the app with different font sizes, we found that the issue "Dynamic Type font sizes are unsupported" was incorrectly reported in two cases for Landmarks app. Given these factors, the tool's effectiveness in practice may be higher than what is reflected in our current report.

We also hypothesized that the capabilities of \name extend beyond the issues reported by the Inspector. In one experiment on a reported issue on GitHub for an iOS app~\cite{ODS_iOS_Issue_703}, we attempted to fix the ``incorrect focus order'' issue using \name. Due to the lack of accessibility identifiers and GUI tests, we manually localized the code snippet and allowed the tool to perform the planning and fixing phases. We provided the screenshot and the title of the report to the model. Given the related code snippet, the model's first plan was to ``set the accessibilityElements property of the parent view.'' The model generated a fix, adding `\texttt{.accessibilityElement(children:.contain)}' to the parent view, which was very similar to the fix submitted by developers. To ensure there was no data leakage to the LLM, we confirmed that the commit date for fixing that issue was after the knowledge cut-off date of GPT-4. While this experiment demonstrates that \name's capabilities can extend beyond the issues reported by the Accessibility Inspector, further experiments are needed.

\subsection{RQ2. \name's efficiency }
In this research question, we assessed \name's efficiency in terms of the number of attempts, time, and number of tokens required to fix issues. 

In terms of the number of attempts, our experiments indicate that out of 363 plausible fix suggestions, \name generated 157 of them on the first attempt, while \name generated 124 and 82 on the second and third attempts, respectively. For the majority of the plausible fix suggestions (57\%), the feedback loop design helped \name resolve the reported failures and generate a plausible fix suggestion, positively impacting the model's effectiveness. However, this improvement in effectiveness comes at the cost of efficiency. By making this feedback loop a configurable parameter, users can adjust it to match their specific resource constraints, thereby balancing efficiency and effectiveness.

We also evaluated the time required for the tool to fix the issues. The mean time to propose alternative fix suggestions (averaged across 10 randomly selected issues) was 54 seconds. Therefore, for a screen with 10 issues, the tool can provide solutions in less than 10 minutes, demonstrating its practical usability. The breakdown of this time is as follows:

The time required for the Data Processing module to statically analyze the app depends on the project's complexity and the number of views it contains. For the apps in our test set, it takes an average of 6 ms to extract the UI hierarchy and instrument the code. Extracting accessibility issues involves building the app, running the GUI test, and using the Accessibility Inspector to dynamically assess the app. These steps, performed by both the Data Processing and Suggestion Assessment modules, take an average of 130 ms per screen.

\name's performance is also closely tied to the LLM response time. We use a publicly available API to communicate with the agents, and various factors, such as online traffic and token constraints per minute, can impact the model's response time. In our experiments, the average response times (across 10 randomly sampled issues) for the Planner, Localizer, and Fixer were 7s, 29.5s, and 8.2s, respectively. The Localizer's longer response time is due to its need to evaluate multiple candidate snippets, requiring more than one inquiry to the model.

In addition to time considerations, the cost of using LLMs is closely related to the number of tokens processed. The average number of tokens per inquiry to the Planner, Localizer, and Fixer is 10K, 15K, and 20K, respectively. The GPT-4o model used in this study costs \$5 per 1M tokens. For a screen with about 10 issues, the total cost of using the tool is less than \$10. 

\subsection{RQ3. \name's helpfulness}
To evaluate the helpfulness of \name in assisting developers with fixing accessibility issues, we conducted a survey of 12 iOS developers within our company. We recruited them from a participant pool from prior studies with around 60 candidates. In the survey,  developers rated suggestions produced by \name and gave feedback on the overall usefulness of the tool. The developers self-rated their SwiftUI iOS development experience on a scale from 1 (No Experience) to 5 (Expert). The developers' median self-rated expertise in SwiftUI app development was 5 (\begin{sparkline}{3}
\definecolor{sparkspikecolor}{gray}{0.9}
    \sparkspike .0 .05
    \sparkspike .2 .05
    
\definecolor{sparkspikecolor}{gray}{0.0}
    \sparkspike .4 .33
    \sparkspike .6 .58
    \sparkspike .8 .82
\end{sparkline}) and in accessibility testing was 5 (\begin{sparkline}{3}
\definecolor{sparkspikecolor}{gray}{0.9}
    \sparkspike .0 .05
\definecolor{sparkspikecolor}{gray}{0.0}
    \sparkspike .2 .25
\definecolor{sparkspikecolor}{gray}{0.9}
    \sparkspike .4 .05
\definecolor{sparkspikecolor}{gray}{0.0}
    \sparkspike .6 .17
    \sparkspike .8 .58
\end{sparkline}). We excluded developers who self-rated as 1 (No Experience) for either of these questions from answering the survey.

\begin{figure}
  \includegraphics[width=\columnwidth]{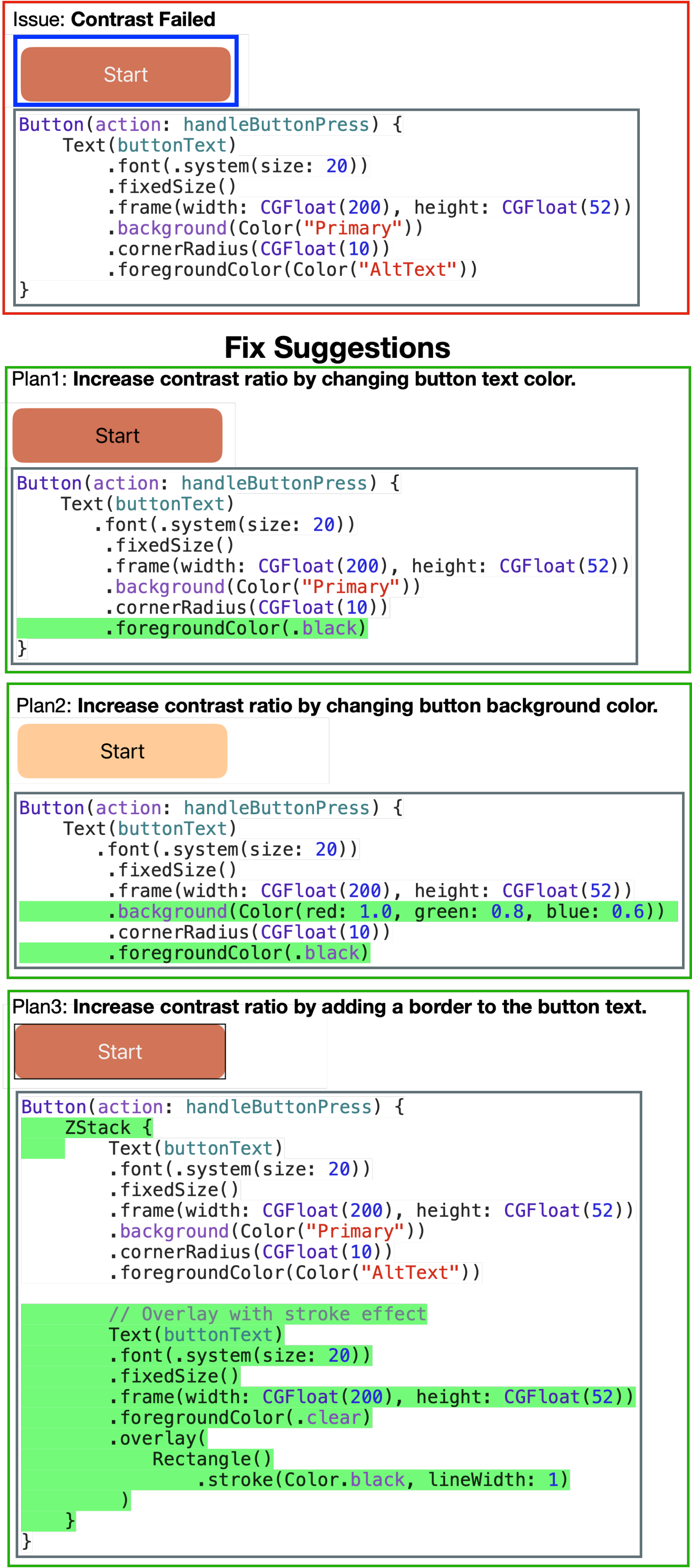}
  \caption{A sample issue with three fix suggestions generated by \name. The first image shows the "Contrast Failed" issue and its original code snippet in Ratio app. The following three boxes represent different plans, each with the modified code snippet and the corresponding screenshot updates.
}
  \label{fig:sample}
\end{figure}

We randomly selected 9 issues and generated plausible fixes for them using \name. \cref{fig:sample} illustrates one of these issues with three alternative code suggestions to fix the issue. For each issue, we showed developers an annotated screenshot with the issue reported by Accessibility Inspector along with a diff visualization for each of the three suggestions and the fix plan and explanation of changes from the LLM. 

\subsubsection{Acceptance of Fix suggestions}
Initially, the survey asked the developers to describe what the issue meant and if they had a hypothesis and a plan for how to fix the issue before showing them the fix suggestions. The developers overall accessibility testing expertise was relatively high. We presented the developers three fix suggestions from the tool and asked whether they would accept any of the proposed suggestions. Overall, developers accepted a suggestion as is or with some modifications for 70\% of the issues. 

Developers did not always accept a fix suggestion that aligned with their initial hypothesis and fix plan. An accessibility testing expert (also an author) assessed ``alignment'' of developers' hypothesis and fix plan matched the corresponding fix suggestion using a scoring rubric of 0 -- did not match, 1 -- matched with some conceptual difference or different level of specificity, and 2 -- perfect match. The mean score for ``alignment'' was 1.1 (Med: 1, $\sigma = 1.31$) and 68\% of the accepted fixes at least partially aligned with the developers' initial fix plan and hypothesis. 

\subsubsection{Helpfulness}
Developers also rated how helpful (from 1 - least helpful to 5 - most helpful) they would find a tool that proposed suggestions for fixing accessibility issues. Developers median rating for helpfulness was 3 (\begin{sparkline}{3}
\definecolor{sparkspikecolor}{gray}{0.9}
    \sparkspike .0 .05
\definecolor{sparkspikecolor}{gray}{0.0}
    \sparkspike .2 .41
    
    \sparkspike .4 .17
    \sparkspike .6 .25
    \sparkspike .8 .17
\end{sparkline}). Less experienced developers in accessibility testing rated the tool as more helpful.

Developers rating the tool a 5 or 4 (6 developers)  found it useful the tool gave multiple suggestions, and noted it was especially useful to help less experienced developers in accessibility testing and implementation to understand different fix strategies. 

While developers who rated the tool a 2 (Slightly Helpful) or 3 (Moderately Helpful) (6 developers) found the \name's localization of the issue helpful, and liked that it provided several options, they critiqued some solutions for not addressing the root of the issue or for being sub-optimal. All developers rating \name's helpfulness as 2 or 3 predominantly focus on accessibility engineering and testing for their job, while the remaining developers (rating helpfulness as 4 or 5) primarily focus on software engineering and occasionally perform accessibility testing. 

The developers' opinions were also mixed on whether it would be useful for the tool to also provide sub-optimal fix suggestions or suggestions which do not fix the issue. Some thought it would be useful to see alternative suggestions to stoke the developer's imagination for finding the right solution (6 developers), while some (3 developers) thought providing these suggestions which were known to not fix the issue could be confusing or distracting, especially to developers less experienced with accessibility testing. 

\subsubsection{Plausible fix vs correct fix.} With \name, our goal was to generate plausible fixes—code modifications that pass the accessibility checks of an automated scanner without introducing new issues or removing any functionality. Our survey shows that out of 36 developer assessments of plausible fix suggestions, only 11 cases did not receive approval for any of the suggested fixes, resulting in a 69.4\% developer acceptance of fixes. However, further studies are needed to understand the various considerations app development teams take into account when determining a correct fix, before automating that process.

\section{Threats to Validity}
\label{Sec:Threats}
\textbf{External Validity:}
One limitation of \name is that it assesses issues individually, even though many issues may be interconnected. For example, fixing one issue might resolve others, or altering the appearance of one element might require adjustments to other related elements. To address this, we have designed the Fixer prompt to cascade design changes across elements to maintain design integrity. As Figure~\ref{fig:results}(c) shows, the model can consider this aspect in some cases. However, without a clear definition of relevant elements and design integrity, the tool's limitations and capabilities are unknown. Future work could focus on developing metrics to assess design integrity or grouping related issues to propose unified solutions and enhance the performance of the tool.

Additionally, \name focuses on single-view files for localizing and generating fixes for issues, which means it cannot address problems that require changes across multiple files. Although the issues in our test set could be resolved within single files, studying more complex, cross-file issues—albeit less common—remains a compelling area for future research. This limitation also impacts the consideration of overall app design integrity beyond a single screen. While providing multiple suggestions allows developers to choose the most suitable one based on app design decisions, incorporating techniques to group issues across different app screens would enhance the tool.

Lastly, the tool has been implemented and tested on iOS apps. We believe that the benefits of \name can be generalized to other platforms by using appropriate tools to build, instrument, and audit apps on those platforms. The system definitions for agents can also be adjusted according to the platform, such as specifying expertise in SwiftUI for iOS or in Android development for Android apps. However, this generalizability needs to be further evaluated. 

\textbf{Internal Validity:}
We implemented \name using various tools and libraries, including XCTest, Accessibility Inspector, and the Tree-sitter library for code parsing~\cite{tree_sitter}. These external tools may introduce defects into the system, and the prototype itself may contain implementation bugs. To mitigate these issues, we tested the tool on a variety of apps at different stages and ensured that we used the latest updates of the external tools.

\section{Related Work}
\label{Sec:Related}
Our work fits into the space of automated accessibility testing and repair tools, which have advanced the state-of-the-art for automated detection and reported of accessibility issue. Our work is among the first to localize and suggest fixes in code for these issues. We also review work LLM-based program repair and fault localization which use similar multi-agent architecture, but do not address GUI or accessibility issues. 

\subsection{Automated accessibility testing and repair}
Many automated tools have been released and proposed in research over the years to detect accessibility issues. Static tools~\cite{googleAndroidLint} examine code directly to find potential issues. One limitation with these tools is that they do not have access to the run-time interface that can be created programmatically or injected with data at runtime. Test-time tools can detect issues from the runtime UI~\cite{salehnamadi2021latte, earlGrey2022, googleEspresso} but are limited by the coverage of the input UI tests which prior work suggests may not exist or have very incomplete coverage over UI states~\cite{padure2020comparing}. Accessibility scanners that examine a run-time interface can detect different classes of issues that surface at run-time~\cite{accessibilityInspector, googleAccessibilityScanner, salehnamadi2021latte} and do not rely on pre-existing UI tests. However, accessibility scanners have two main limitations: 1) they do not assist developers in effectively localizing the impacted UI element with an issue, and 2) they provide developers with little assistance in fixing the issue other than sometimes a single high-level non-contextualized fix suggestion. \name takes as input an issue detected by one of these tools, its description, information about the impacted UI element in the form of a screenshot  (i.e., Accessibility Inspector~\cite{accessibilityInspector}), and the app source code, and both localizes the UI element and provides multiple relevant fix suggestions. 

Some work combines run-time accessibility scanners with app crawlers to detect and report accessibility issues~\cite{salehnamadi2022groundhog, eler2018automated, swearngin2023towards}. While these tools can surface more issues to developers, the amount of issues reported by these tools can be overwhelming~\cite{huq2023a11ydev} especially when even localizing and fixing one issue is already challenging. These tools are also not connected to the underlying source code, which can often be a reason for developers to ignore the reported issues from these tools. 

Another area of work has developed single-purpose, mostly machine-learning based, techniques to detect specific accessibility issues such as color issues~\cite{zhang2023automated}, touch target size~\cite{alotaibi2021automated}, missing labels~\cite{mehralian2021data}, and text scaling~\cite{alotaibi23icsme}. These solutions have predominately focused on Android apps, which have very different specification than iOS apps and are not implemented in SwiftUI.  Zhang et. al~\cite{zhang2021screen} detect and repair UI elements for the iOS VoiceOver screen reader using an approach that could be platform agnostic. However, there remains a huge gap between these solutions and the original source code where the developer must make the fix. Even if the localization issue were solved by these methods, there would still be a need to generate alternate fix solutions. As we learned in our formative interviews, fixing one issue may introduce other issues, so there is likely no one-size-fits-all fix for each issue category. Developers need to consider many other requirements to determine the correct fix (e.g., input from design teams, impact on other areas of the app). In contrast to prior solutions, \name both localizes the code for the impacted UI element and suggests multiple candidate fixes. Future versions of \name can also incorporate some of these techniques into its pipeline for issue detection, and then rely on the capabilities of its LLM to suggest candidate solutions.

\subsection{LLM-based program repair and fault localization}
Recent progress in LLMs have advanced their application in automating program repair tasks. A systematic survey on LLMs for automated program repair~\cite{zhang2024systematic} reviewed 127 studies, covering various aspects of this problem, but none specifically address GUI or accessibility issues.

The only study focused on GUI issues is ACCESS~\cite{huang2024access}, which examines LLM capabilities in correcting web accessibility violations by exploring different prompt engineering techniques to fix issues reported on specific HTML tags. This approach is limited to textual data from web pages and does not address mobile app issues, where it is necessary to identify the location of the problematic GUI element in the code.

Recently, researchers have expanded beyond prompt engineering to enhance LLM-based bug repair performance for GitHub issues. They have employed various techniques, including fine-tuning on specific datasets \cite{wang2023rap, BerkayBerabi2021, Fu2022, Mashhadi2021, Lajko2022}, or more advanced strategies such as retrieval-augmented generation to guide search space~\cite{jimenez2023swe}, agents interacting with other tools~\cite{yang2024swe, bouzenia2024repairagent}, or multi-agent systems operating in different steps~\cite{zhang2024autocoderover, tao2024magis, chen2024coder}.

In these works~\cite{chen2024coder, tao2024magis}, researchers drew inspiration from human roles in real-world scenarios, such as Manager and Developer, to design LLM agents with specific actions. Tao et al.~\cite{tao2024magis} enforce collaboration between agents to localize a file and implement a patch. Chen et al.~\cite{chen2024coder} proposed four collaborative plans for agents to address issues, with the Manager agent selecting a plan from predefined options. These studies demonstrate how modeling software engineering processes with agents can enhance issue resolution capabilities. However, these approaches only address functional issues reported on GitHub.

In contrast to prior related work, our work targets accessibility issues in mobile apps using a novel multi-agent LLM architecture. It designs agents inspired by the steps developers take to solve these problems while also considering their unique characteristics, such as the diversity of accessibility issues, various fix strategies, and the need to analyze GUI images with reported issues. We developed \name, which analyzes issues in mobile apps, employs a plan-localize-fix process, and produces plausible fixes. Additionally, \name incorporates a feedback loop that uses natural language error messages to enable the LLM to reflect~\cite{shinn2024reflexion} on and address its own failures.

\section{Conclusion}
\label{Sec:Conclusion}
Fixing accessibility issues in mobile apps is a challenging task for developers. While automated scanners can identify these issues, they often fall short in guiding developers to the exact location in the code and suggesting appropriate fixes. Towards addressing these needs, we proposed a plan-localize-fix technique, operationalized through a multi-agent LLM architecture called \name. Evaluations on \name demonstrate its capabilities in suggesting plausible fix suggestions and highlight how the tool assists the developer in the decision-making process for selecting and applying accessibility-related fixes. Our results suggest that applying LLMs to fix accessibility issues in source code is an encouraging direction.

\bibliography{references}
\bibliographystyle{IEEEtran}

\end{document}